\documentclass[12pt]{iopart}
\pdfoutput=1
\usepackage{graphicx}

\begin{document}
\title{Mechanical memory for photons with orbital angular momentum}

\author{H. Shi and M. Bhattacharya}

\address{School of Physics and Astronomy, Rochester Institute of Technology, 84 Lomb Memorial Drive, Rochester, NY 14623, USA}
\ead{mxbsps@rit.edu}
\begin{abstract}
We propose to use an acoustic surface wave as a memory for a photon carrying orbital angular
momentum. We clarify the physical mechanism that enables the transfer of information, derive the
angular momentum selection rule that must be obeyed in the process, and show how to optimize the
optoacoustic coupling. We theoretically demonstrate that high fidelities can be achieved, using
realistic parameters, for the transfer of a coherent optical Laguerre-Gaussian state, associated
with large angular momentum, to a mechanical shear mode. Our results add a significant possibility
to the  ongoing efforts towards the implementation of quantum information processing using photonic
orbital angular momentum.
\end{abstract}
\vspace{2pc}

Significant developments in contemporary optics have resulted from the realization that photons
can carry orbital angular momentum (OAM) encoded in the spatial distribution of the corresponding
electromagnetic field \cite{OAMPRA1992,OAMBook}. Distinct from the photon polarization, which can provide a maximum
angular momentum of $\hbar$, the field profile can be engineered to endow the photon with virtually unlimited
OAM, $l\hbar$. Recent experiments have produced up to $l=300$ \cite{Zeilinger2012}. For high $l$, each photon
can carry a large amount of information in a multi-dimensional Hilbert space. This capability makes
photonic OAM states attractive for quantum information processing tasks
\cite{Mair2001,Terriza2002,Zou2005,Giovannini2011,Escartin2012}
some of which are more efficiently \cite{Lanyon2009}, stably \cite{Collins2002} or
securely \cite{Cerf2002} accomplished in higher dimensional spaces.

Photons carrying OAM constitute the ``flying" qudits of a quantum information processing scheme, as they can
reliably convey information between remote locations. In addition, ``memories" are also required, to store
the information at those locations \cite{WalmsleyReview}. With regard to storing OAM, theoretical proposals
\cite{Marzlin1997,Dum1998,Nandi2004,Dutton2004,Kapale2005,Kanamoto2007,Thanvanthri2008,Paternostro2010} as well as
experiments \cite{Tabosa1999,Inoue2006,Andersen2006,Pugatch2007,Wright2008,Moretti2009,Inoue2009,Wright2009,
Leslie2009,Guo2006,Guo2008} have focused on atomic media, such as vapors and degenerate gases. However,
atomic experiments provide complex, somewhat fragile, and difficult-to-scale platforms for photonic OAM memory.

In this article we propose to store OAM states in a simpler, more robust, and readily scalable memory: acoustic
modes on the surface of an optical mirror \cite{Briant2003}. When the mirror is part of a Fabry-Perot cavity, the
acoustic modes can phase-modulate an intracavity optical mode in a manner identical to a global displacement of
the mirror. The optoacoustic interaction responsible for this phase modulation can be traced back to radiation
pressure similar to the many optomechanical systems being realized currently
\cite{VahalaReview,GirvinReview,AspelmeyerReview}. Extending the analogy further, we suggest that this optoacoustic
coupling can be used to store and retrieve optical information from a mechanical degree of freedom, as demonstrated
recently for optomechanical systems \cite{Tian2011}. We note that earlier optomechanical proposals involving
optical OAM have relied on torsional
oscillators \cite{Bhattacharya2007,Isart2010,Shi2013}, which, not being free rotors, offer states rather localized
in angle, which provide only small spatial overlap with optical OAM states. In the present case however, as we
show below, a high degree of mode-matching can be achieved between radiation and matter, leading to a
large coupling and good memory fidelity.

\begin{figure}[htbp]
\begin{center}
\includegraphics[width=0.5\columnwidth]{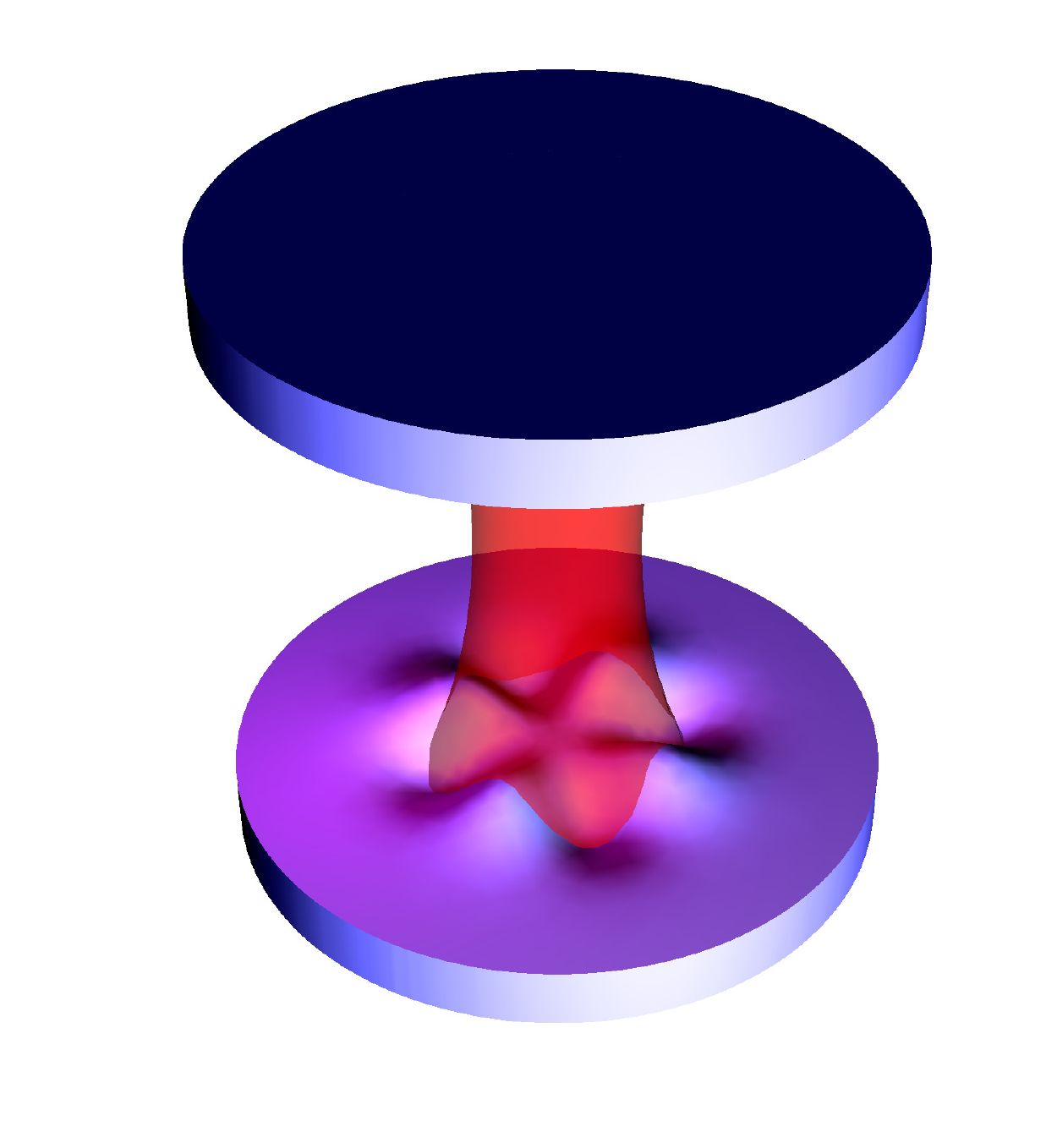}
\caption{An optoacoustical system composed of a two-mirror cavity and counter-propagating and counter-rotating
Laguerre-Gaussian modes. The lower mirror supports surface acoustic waves \cite{Briant2003}.}
\label{fig1}
\end{center}
\end{figure}

Our specific system of interest is shown in Fig.~\ref{fig1}, and consists simply of a two-mirror cavity.
We consider surface acoustic modes on one of the mirrors, as observed in an earlier experiment \cite{Briant2003}. The
acoustic modes consist of mechanical surface vibrations along the optical axis; however these vibrations carry an
azimuthal modulation, i.e. a shear, which can accommodate the photonic OAM, as we will show below. For vibrational
displacement amplitudes much smaller than the mirror thickness, the acoustic modes are well described by the
normalized mode function
\cite{OAMBook,Briant2003,Loudon2003}
\begin{equation}
\fl u_{lp}(r,\theta)=\sqrt{\frac{4p!}{(1+\delta_{0l})\pi(|l|+p)!}}\frac{1}{w}
\left(\frac{\sqrt{2}r}{w}\right)^{|l|}L_p^{|l|}\left(\frac{2r^2}{w^2}\right)\exp{\left(-\frac{r^2}{w^2}\right)}\cos(l\theta),
\label{eq1}
\end{equation}
where the index $l$ corresponds to the number of azimuthal nodes, the index $p$ determines
the number of radial nodes, and $w$ equals the acoustic beam waist $w_{a}$. In Eq.~(\ref{eq1}), the index $l$ can be
either a positive or negative integer, while the index $p$ can only assume positive integer values \cite{OAMBook}.
The intensity of an optical cavity field consisting of the superposition of two counter-rotating Laguerre-Gaussian modes
can also be described in terms of Eq.~(\ref{eq1}), i.e. as $|u_{lp}(r,\theta)|^{2}$, where $l,p$ and $w=w_{c}$
now describe the optical parameters. Below we will consider the transfer of this optical counter-rotating superposition
state $|+l\rangle+|-l\rangle$ \cite{Kapale2005} to a shear mechanical mode \cite{Briant2003}.

We expand the quantized electromagnetic field using the modes of Eq.~(\ref{eq1}) \cite{Loudon2003} with unprimed
indices $l$ and $p$ and similarly the acoustic displacement \cite{Briant2003} but using mode functions with primed indices
$l'$ and $p'$. We find the optoacoustic interaction Hamiltonian to be of the form of the standard `linearly' coupled
optomechanical system \cite{Briant2003,AspelmeyerReview}
\begin{equation}
\label{eq:HInt}
\hat{H}=-i\hbar g\chi_{ll'pp'}\hat{a}^\dagger \hat{a}(\hat{b}+\hat{b}^\dagger),
\end{equation}
where $\hat{a}$ ($\hat{a}^\dagger$) and $\hat{b}$ ($\hat{b}^\dagger$) are the annihilation (creation) operators for
the optical and acoustical modes respectively. The constant $g$ is given by
\begin{equation}
g=\frac{x_0\omega_{c}}{L},
\end{equation}
where $L$ is the optical cavity length, $x_0$ is the zero-point-motion amplitude of the surface acoustic mode, and
$\omega_c$ is the frequency of the free cavity field. The dimensionless
constant $\chi_{ll'pp'}$, which is the overlap integral between the optical and acoustical modes,
\begin{equation}
\label{eq:OL}
\chi_{ll'pp'}=w_{a}\int_{r=0}^{\infty} \int_{\theta=0}^{2\pi}|u_{lp}(r,\theta)|^{2}u_{l'p'}(r,\theta)d\theta r d r,
\end{equation}
is analytically found to be
\begin{equation}
\chi_{ll'pp'}=\xi_{lpp'}\delta_{|l'|,2|l|},
\end{equation}
where
\begin{eqnarray}
\fl\xi_{lpp'}=&&\frac{p!}{(1+\delta_{0,l})(|l|+p)!}\sqrt{\frac{p'!}{(1+\delta_{0,2l})\pi(2|l|+p')!}}\nonumber\\
\fl&&\times\frac{\Gamma(p+|l|+1)}{2^{2p}p!}\sum_{k=0}^p\left(\begin{array}{c}2p-2k\\ p-k\end{array}\right)\frac{\Gamma(p'+2k+2|l|+1)}{k!p'!\Gamma(|l|+k+1)}\frac{(1-\gamma/2)^{2k+p'}}{(1+\gamma/2)^{2k+p'+2|l|+1}}\nonumber\\
\fl&&\times   {}_{2}F_{1}\left[-p',-2k;-p'-2k-2|l|;\left(\frac{1+\gamma/2}{1-\gamma/2}\right)^2\right],
\label{eq5}
\end{eqnarray}
with
\begin{equation}
\gamma=\left(\frac{w_{c}}{w_{a}}\right)^2.
\end{equation}
In Eq.~(\ref{eq5}), $\Gamma(p+|l|+1)$
is a Gamma function, and $_{2}F_{1}$ is a hypergeometric function. We note that $\chi_{ll'pp'}$ is only non-zero if the angular
momentum conservation rule
\begin{equation}
\label{eq:SR}
|l'|=2|l|,
\end{equation}
is satisfied. The factor of two can be traced to the quadratic dependence of the integral of Eq.~(\ref{eq:OL}) on the
optical mode function as compared to its linear dependence on the acoustic mode, and even further back to the bilinear
dependence of the interaction on the optical operators and its linear dependence on the acoustic displacement, see
Eq.~(\ref{eq:HInt}). Henceforth, we will assume that Eq.~(\ref{eq:SR}) is satisfied and will deal only with the quantity
$\xi_{lpp'}$. Similarly, we will consider $w_{a},$ which is determined by the mirror geometry \cite{Briant2003}, to be a
fixed parameter, and consider variations only in $w_{c}$, or equivalently in $\gamma$, below.

We now first examine the simple case $p=p'=0$, in which case Eq.~(\ref{eq5}) reduces to
\begin{equation}
\xi_{l00}=\frac{\gamma^{|l|}}{(1+\delta_{0,l})|l|!}\sqrt{\frac{(2|l|)!}{(1+\delta_{0,2l})\pi}}\left(1+\frac{\gamma}{2}\right)^{-(2|l|+1)}.
\end{equation}
The value of $\gamma$ that maximizes the overlap $\xi_{l00}$ is given by
\begin{equation}
\gamma_{max}=\frac{2|l|}{|l|+1},
\end{equation}
for $|l| \neq 0.$ It can be readily verified that the corresponding $\xi_{l00}$ reaches a maximum value of $0.23$ for
the case of lowest nontrivial OAM $|l|=1$ and decreases monotonically for larger $l.$

\begin{figure}[htbp]
\begin{center}
\includegraphics[width=0.5\columnwidth]{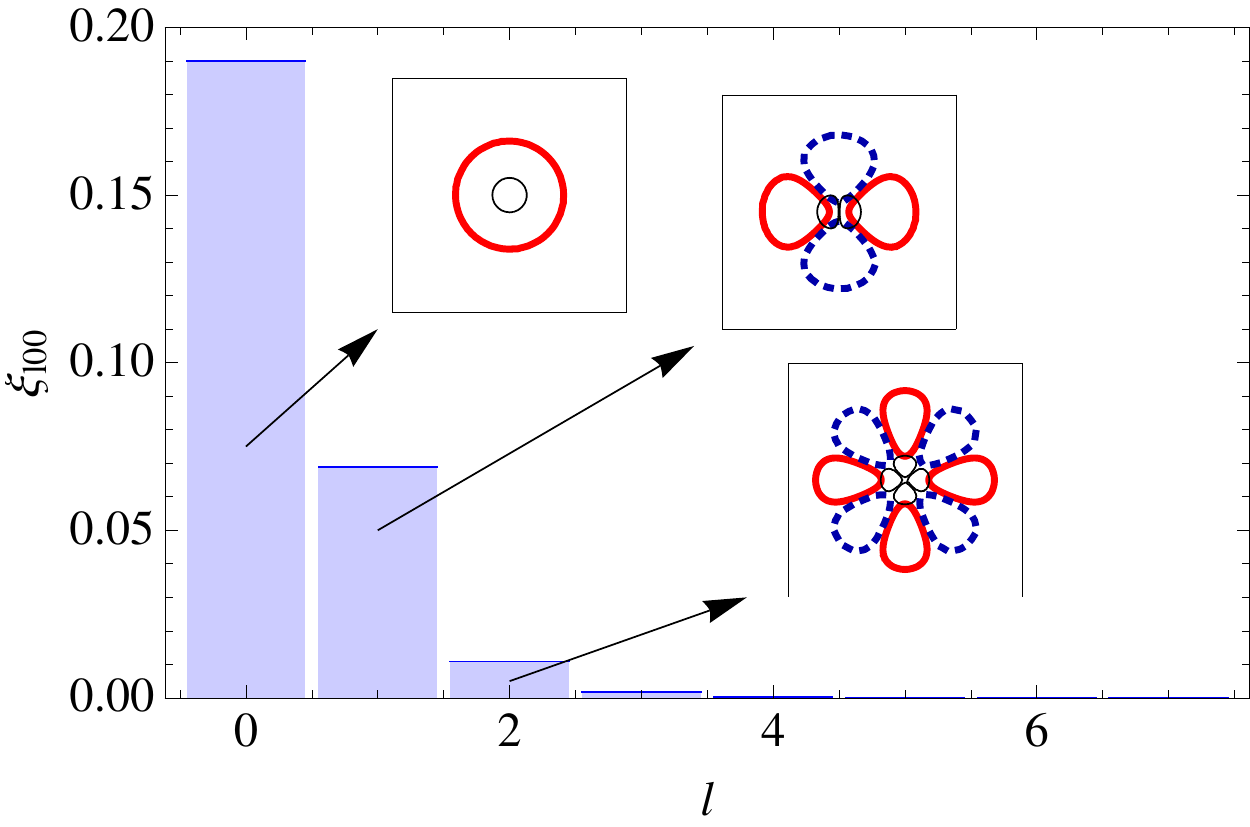}
\caption{Dimensionless coupling constant $\xi_{lpp'}$ for the case $p=p'=0$ with $\gamma=0.1$. The three insets
shows the corresponding contours of the optical mode profile (thin solid lines) and the mechanical mode (thick solid line
for positive  displacements and thick dotted line for negative displacements). The coupling constant decreases rapidly as
a function of $l$ due to the decreasing overlap between the optical and acoustic modes.}
\label{fig2}
\end{center}
\end{figure}

In some situations, however, it may be desirable to have control of $\gamma$ independent of $|l|.$ For example, making
$\gamma$ small reduces the effective mass of the acoustic mode and enhances quantum effects \cite{Pinard1999}.
In this case of sub-optimal $\gamma \neq \gamma_{max}$, $\xi_{l00}$ also decreases with $l$
as shown in Fig.~\ref{fig2} for $\gamma=0.1$. The negligible coupling between the optical mode and the
acoustic mode for large values of $l$ prevents the exploitation of the in-principle infinite dimensional Hilbert
space spanned by OAM photonic states. We will now show that  non-zero values of $p$ \cite{Ueda2010} or $p'$
\cite{Briant2003}, can be used to generate substantial couplings for large values of $l$.

\begin{figure}[htbp]
\begin{center}
\includegraphics[width=0.75\columnwidth]{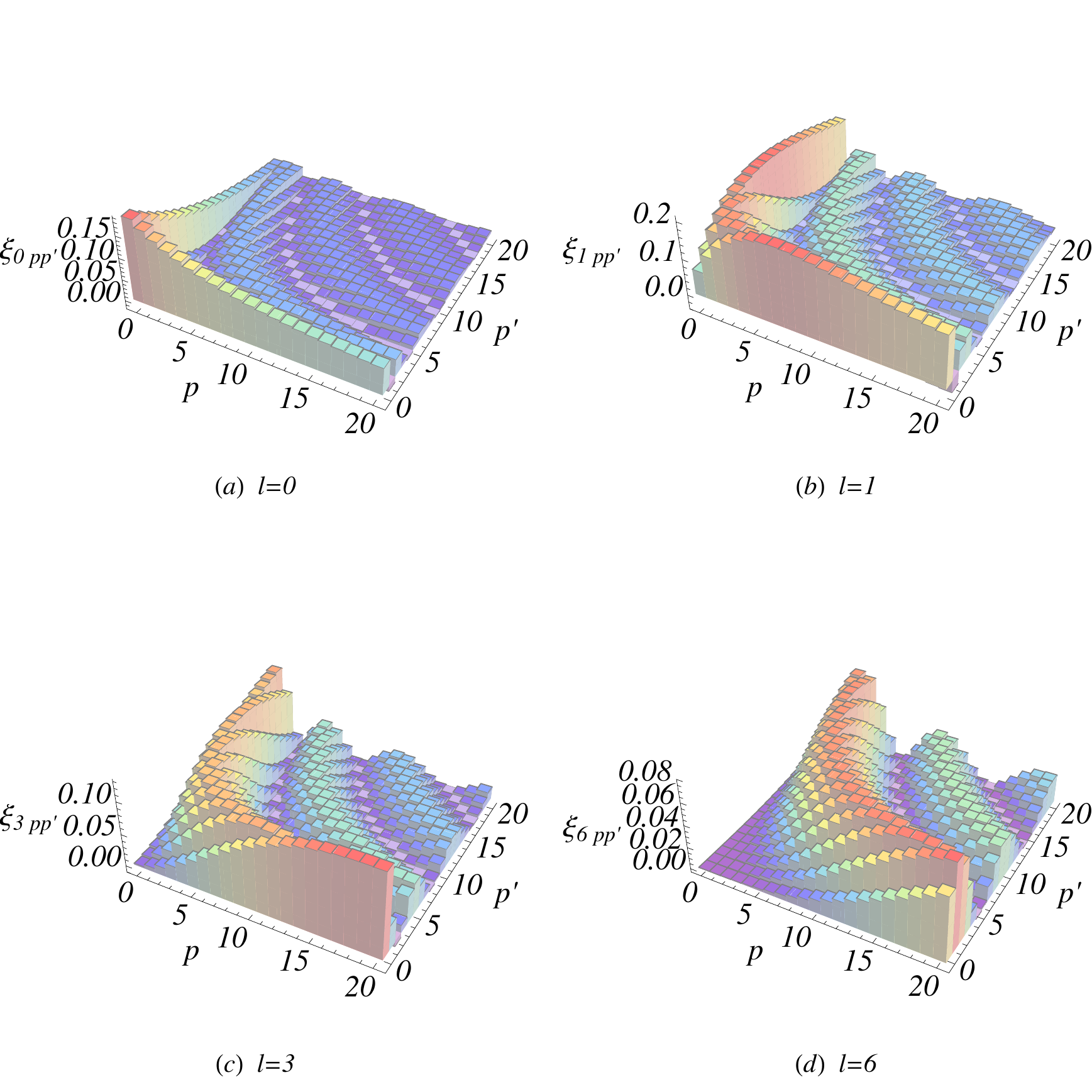}
\caption{Dimensionless coupling constant $\xi_{lpp'}$ as a function of $p$ and $p'$ for $\gamma=0.1$ and (a) $l=0$,
(b) $l=1$, (c) $l=3$, and (d) $l=6$. The use of non-zero $p$ and $p'$ enables the generation of significant
coupling for optical states with high values of $l$.}
\label{fig3}
\end{center}
\end{figure}

\begin{figure}[htbp]
\begin{center}
\includegraphics[width=0.5\columnwidth]{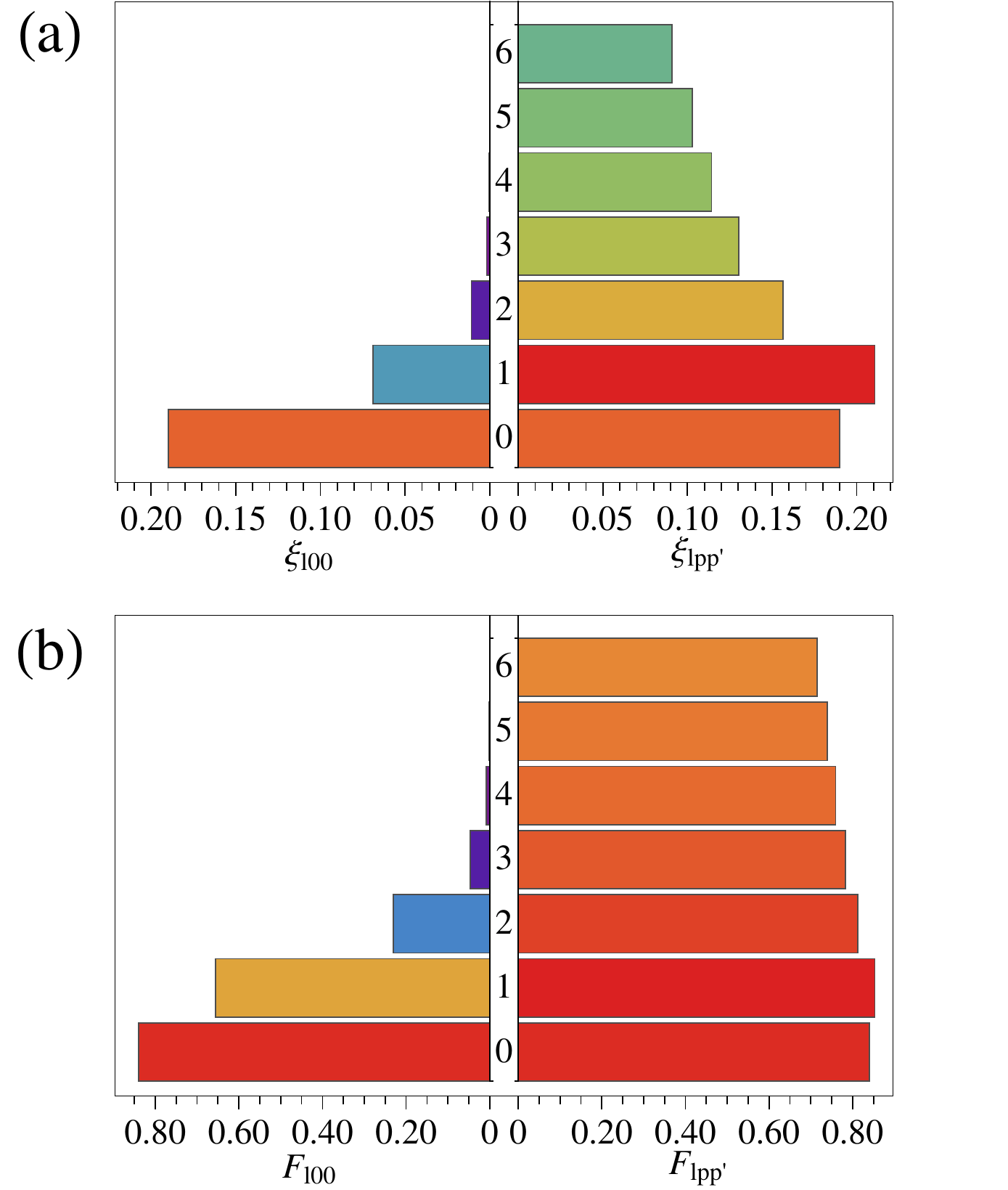}
\caption{Comparison of the (a) dimensionless coupling constant $\xi_{lpp'}$ and (b) state transfer fidelity with $p=p'=0$ (left panel) and with $p\neq0$ and
$p'\neq0$ (right panel), with $\gamma=0.1$. The right panel is generated by finding the maximum coupling (fidelity) that can be achieved for a
given value of $l$ without restrictions on $p$ and $p'$. It can be seen that the use of non-zero values of $p$ and $p'$
makes it possible to achieve significant optoacoustic coupling and high state transfer fidelity for high values of $l$.}
\label{fig4}
\end{center}
\end{figure}

Before proceeding, we note that the effect of non-zero values of $p (p')$ on the mode profile
is two fold. First, non-zero values of $p (p')$ introduce additional local maxima in the radial
direction. This means that the intensity distributed to the central peaks decreases as $p(p')$
become larger, for the same value of total intensity. Secondly, the radius at which the central intensity peaks
occur is reduced. In Fig.~\ref{fig3}, we have shown $\xi_{lpp'}$ as a function
of $p$ and $p'$ for different values of $l$, with $\gamma=0.1$.
For $l=0$, the optimal arrangement of the optical and acoustical mode
profiles occurs at $p=p'=0$, where the two peaks are exactly aligned. However, for $l>0$, non-zero values of $p$ or $p'$
help to increase the coupling constant to about the same order of magnitude as that in the case $l=p=p'=0$. In
Fig.~\ref{fig4}(a), a comparison of the coupling strengths between the cases of $p=p'=0$ (left) and non-zero values of $p$ and $p'$
(right) has been exhibited. As can be seen, even though the coupling constant still decreases as $l$ gets larger, the rate
of decrease is much slower than that in Fig.~\ref{fig2}. In fact, for quite high OAM, $\xi_{lpp'}$ can be adjusted to be
of the same order of magnitude as that for $l=0$.

\begin{figure}[htbp]
\begin{center}
\includegraphics[width=0.75\columnwidth]{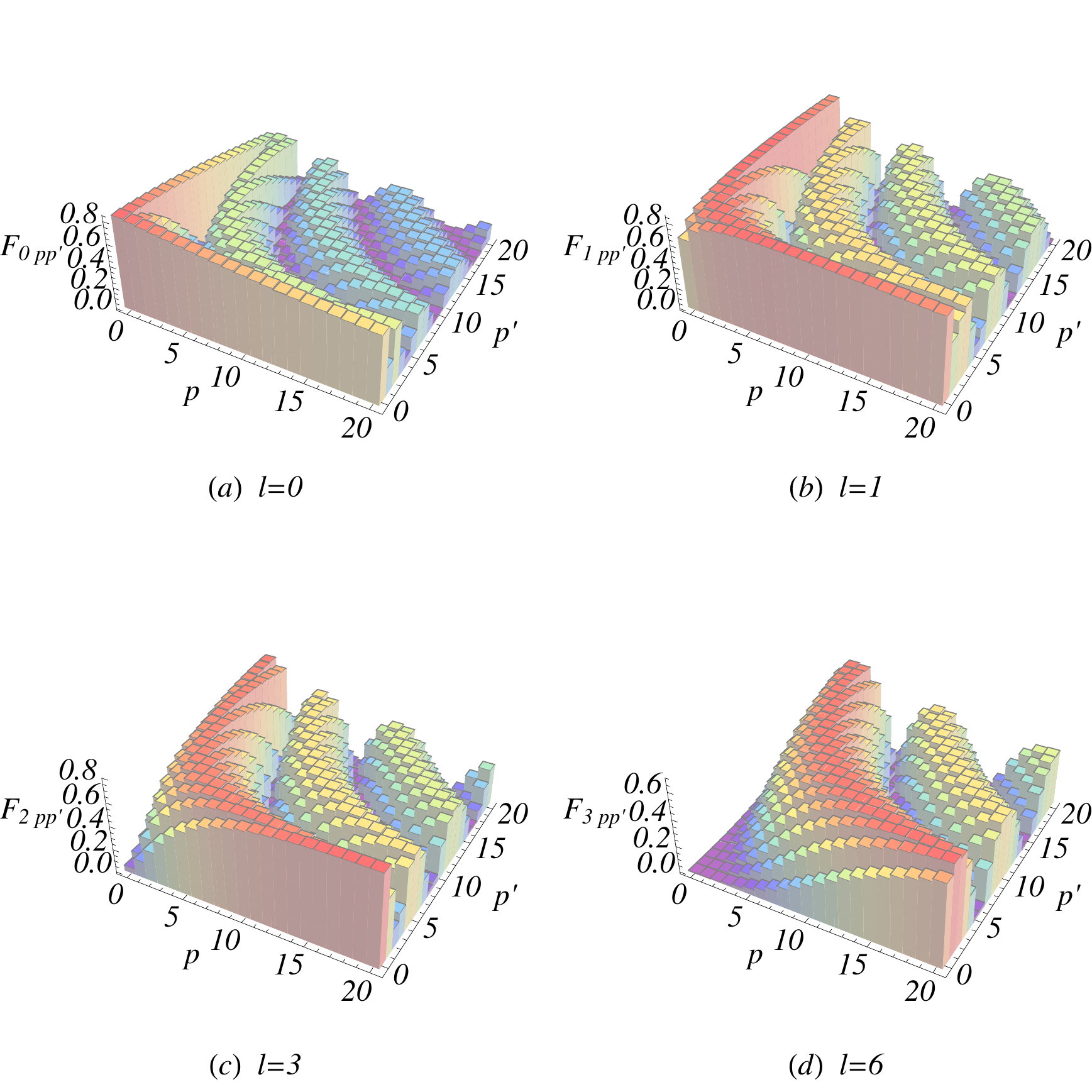}
\caption{Fidelity for the transfer of a coherent optical state $|\alpha\rangle$ to the mechanical mode, with
$g=2\pi\times 0.2$Hz, $n_{c}=2\times 10^{18}$, $\gamma=0.1$, $\alpha=1$, $\kappa=2\pi\times50$kHz, $\gamma_m=2\pi\times50$kHz,
and $T=1$K \cite{Clerk2012}.}
\label{fig5}
\end{center}
\end{figure}

We have also calculated the fidelity for transferring a coherent optical state $|\alpha\rangle$ in the OAM superposition
$|+l\rangle+|-l\rangle$ to the mechanical shear mode, accounting for the effects of noise and dissipation on both degrees
of freedom. We have adapted an earlier treatment to calculate the fidelity
of the transfer as a function of the indices $l$, $p$, and $p'$ \cite{Clerk2012}. Assuming a strongly driven optical cavity
and small single photon optomechanical coupling $g$, the fidelity for the transfer of a coherent state is given by
\begin{equation}
F_{lpp'}=\frac{1}{1+n_{lpp'}}\exp\left(-\frac{\lambda_{lpp'}^2}{1+n_{lpp'}}\right),
\end{equation}
where
\begin{equation}
n_{lpp'}=\frac{\gamma_m N_m}{2}\frac{\pi}{2g\sqrt{n_{c}}\xi_{lpp'}}
\label{eq8}
\end{equation}
is the effective number of thermal quanta responsible for heating the state during transfer and
\begin{equation}
\lambda_{lpp'}=\alpha \left(\frac{\kappa+\gamma_m}{4}\right)\frac{\pi}{2g\sqrt{n_{c}}\xi_{lpp'}},
\label{eq9}
\end{equation}
is a damping parameter accounting for dissipative effects. In Eqs.~(\ref{eq8}) and (\ref{eq9}), $\gamma_m$ is the
acoustic mode decay rate, $\kappa$ is the optical cavity decay rate, $N_m$ is the number of phonons in the environment,
$\alpha$ is the coherent amplitude of the state to be transferred and $n_{c}$ is the number of intracavity photons due to
the optical drive. We have chosen experimentally relevant parameters for our calculation, with a cavity decay rate of
$\kappa=2\pi \times 50$kHz, a mechanical decay rate of $\gamma_m=2\pi \times 50$kHz, a drive of power $\sim 1$mW ($n_{c}\sim 10^{18}$) 
and an environmental temperature of $T=20$mK \cite{Briant2003,Clerk2012,AspelmeyerReview}. The fidelity for the transfer of a
coherent state with $\alpha=1$ has been plotted in Fig.~\ref{fig5} for a few values of $l$.
The transfer fidelity for high photon angular momentum ($l\geq3$) is quite low with $p=p'=0$. However, the use of non-zero
values of $p$ and $p'$ can restore the transfer fidelity for states with high values of $l$ to the same level that can only
be achieved for $l=0$ with $p=p'=0$. A comparison for the state transfer fidelity between the case of $p=p'=0$ and $p\neq0,~p'\neq0$ is shown in Fig. \ref{fig4}(b).

In conclusion, we have demonstrated that substantial optomechanical coupling can be achieved in a system composed of an
acoustical mode and a Laguerre-Gaussian cavity mode, provided that ratio of the acoustic and optical waists as well as the
indices $p$ and $p'$ for the modes are chosen judiciously. We have also shown that high fidelity transfer of a coherent
optical state to the mechanical mode can be carried out for experimentally realistic parameters. Our results make it possible
to exploit the in-principle infinite dimensional Hilbert space spanned by the optoacoustic system, and realize a mechanical
memory for photons carrying orbital angular momentum. Unlike previous work, which has focused on atomic vapors, the mechanical
platform we have suggested can easily be miniaturized and readily scaled. We would like to thank the Research Corporation for
Science Advancement for Support.\\
\textbf{References}

\vspace{0.5in}
\begin{thebibliography}{00}

\bibitem{OAMPRA1992} Allen L, Beijersbergen M W, Spreeuw R J C, and Woerdman J P, 1992 \textit{Phys. Rev. A} \textbf{45} 8185

\bibitem{OAMBook} Allen L, Barnett S M, and Padgett M J, Editors, 2003 \textit{Optical Angular Momentum},
(United States: Institute of Physics Publishing)

\bibitem{Zeilinger2012} Fickler R, Lapkiewicz R, Plick W N, Krenn M, Schaeff C, Ramelow S and Zeilinger A,
2012 \textit{Science} \textbf{338} 640

\bibitem{Mair2001} Mair A, Vaziri A, Weihs G and Zeilinger A, 2001 \textit{Nature} \textbf{412} 313

\bibitem{Terriza2002} Molina-Terriza G, Torres J P and Torner L, 2001 \textit{Phys. Rev. Lett.} \textbf{88} 013601

\bibitem{Zou2005} Zou X and Mathis W, 2005 \textit{Phys. Rev. A} \textbf{71} 042324

\bibitem{Giovannini2011} Giovannini D, Nagali E, Marucci L and Sciarrino F, 2011 \textit{Phys. Rev. A} \textbf{83} 042338

\bibitem{Escartin2012} Garcia-Escartin J C and Chamorro-Posada P, 2012 \textit{Phys. Rev. A} \textbf{86} 032334

\bibitem{Lanyon2009} Lanyon B P, Barbieri M, Almeida M P, Jennewein T, Ralph T C, Resch K J, Pryde G. J.,
O'Brien J L, Gilchrist A and White A G, 2009 \textit{Nature Phys.} \textbf{7} 134

\bibitem{Collins2002} Collins D, Gisin N, Linden N, Massar S and Popescu S, 2002 \textit{Phys. Rev. Lett.}
\textbf{88} 040404

\bibitem{Cerf2002} Cerf N, Bourennane M, Karlsson A and Gisin N, 2002 \textit{Phys. Rev. Lett.} \textbf{88} 127902

\bibitem{WalmsleyReview} Simon C et al., 2010 \textit{Eur.Phys. J. D} \textbf{58} 1

\bibitem{Marzlin1997} Marzlin K P, Zhang W and Wright E M 1997 \textit{Phys. Rev. Lett.} \textbf{79} 4728

\bibitem{Dum1998} Dum R, Cirac J I, Lewenstein M and Zoller P 1998 \textit{Phys. Rev. Lett.} \textbf{780} 2972

\bibitem{Nandi2004} Nandi G, Walser R and Schleich W P 2004 \textit{Phys. Rev. A} \textbf{69} 063606

\bibitem{Dutton2004} Dutton Z and Ruostekoski J 2004 \textit{Phys. Rev. Lett.} \textbf{93} 193602

\bibitem{Kapale2005} Kapale K T and Dowling J P 2005 \textit{Phys. Rev. Lett.} \textbf{95} 173601

\bibitem{Kanamoto2007} Kanamoto R, Wright E M and Meystre P 2007 \textit{Phys. Rev. A} \textbf{75} 063623

\bibitem{Thanvanthri2008} Thanvanthri S, Kapale K T and Dowling J P 2008 \textit{Phys. Rev. A} \textbf{77} 053825

\bibitem{Paternostro2010} Gullo N L, McEndoo S, Busch T and Paternostro M 2010 \textit{Phys. Rev. A} \textbf{81} 053625

\bibitem{Tabosa1999} Tabosa J W R and Petrov D V 1999 \textit{Phys. Rev. Lett.} \textbf{83} 4967

\bibitem{Inoue2006} Inoue R, Kanai N, Yonehara T, Miyamoto Y, Koashi M and Kozuma M 2006 \textit{Phys. Rev. A}
\textbf{74} 053809

\bibitem{Andersen2006} Andersen M, Ryu C, Clade P, Natarajan V, Vaziri A, Helmerson K and Phillips W D
2006 \textit{Phys. Rev. Lett.} \textbf{97} 170406

\bibitem{Pugatch2007} Pugatch R, Shuker M, Firstenberg O, Ron A and Davidson N 2007
\textit{Phys. Rev. Lett.} \textbf{98} 203601

\bibitem{Wright2008} Wright K C, Leslie L S and N P Bigelow 2008 \textit{Phys. Rev. A} \textbf{77} 041601(R)

\bibitem{Moretti2009} Moretti D, Felinto D and Tabosa J W R 2009 \textit{Phys. Rev. A} \textbf{79} 023825

\bibitem{Inoue2009} Inoue R, Yonhara T, Miyamoto Y, Koashi M and Kozuma M \textit{Phys. Rev. Lett.}
\textbf{103} 110503

\bibitem{Wright2009} Wright K C, Leslie L S, Hansen A and Bigelow N P \textit{Phys. Rev. Lett.}
\textbf{102} 030405

\bibitem{Leslie2009} Leslie L S, Hansen A, Wright K C, Deutsch B M and Bigelow N P \textit{Phys. Rev. Lett.}
\textbf{103} 250401

\bibitem{Guo2006} Jiang W, Chen Q F, Zhang Y S and Guo G C 2006 \textit{Phys. Rev. A} \textbf{74} 043811

\bibitem{Guo2008} Chen Q F, Shi B S, Zhang Y S and Guo G C 2008 \textit{Phys. Rev. A} \textbf{78} 053810

\bibitem{Briant2003} Briant T, Cohadon P F, Heidmann A And Pinard M 2003 \textit{Phys. Rev. A} \textbf{68} 033823

\bibitem{VahalaReview} Kippenberg T J and Vahala K 2008 \textit{Science} \textbf{321} 1172

\bibitem{GirvinReview} Marquardt F and Girvin S 2009 \textit{Physics} \textbf{2} 40

\bibitem{AspelmeyerReview} Aspelmeyer M, Kippenberg T J and Marquardt F 2013 arXiv:1303.0733 [cond-mat.mes-hall]

\bibitem{Tian2011} Fiore V, Yang Y, Kuzyk M C, Barbour R, Tian L, and Wang H 2011 \textit{Phys. Rev. Lett.}
\textbf{107} 133601

\bibitem{Bhattacharya2007} Bhattacharya M and Meystre P 2007 \textit{Phys. Rev. Lett.}
\textbf{99} 153603

\bibitem{Isart2010} Romero-Isart O, Juan M L, Quidant R and Cirac J I 2010 \textit{N. J. Phys.}
\textbf{12} 033015

\bibitem{Shi2013} Shi H and Bhattacharya M 2013 \textit{J. Mod. Opt.}
\textbf{5} 382

\bibitem{Loudon2003} Loudon R 2003 \textit{Phys. Rev. A}
\textbf{68} 013806

\bibitem{Pinard1999} Pinard M, Hadjar Y and Heidmann A 1999 \textit{Eur. Phys. J. D}
\textbf{7} 107

\bibitem{Ueda2010} Thirugnanasambandam M P, Senatsky Y and Ueda K 2010 \textit{Laser Phys. Lett.}
\textbf{7} 637

\bibitem{Clerk2012} Wang Y D and Clerk A 2012 \textit{N. J. Phys.}
\textbf{14} 105010
\end{thebibliography}

\end{document}